\documentclass{aa}
\usepackage{psfig}
\def\ltsima{$\; \buildrel < \over \sim \;$}
\def\lsim{\lower.5ex\hbox{\ltsima}}
\def\gtsima{$\; \buildrel > \over \sim \;$}
\def\gsim{\lower.5ex\hbox{\gtsima}}
\newcommand{\be}{\begin{equation}}
\newcommand{\en}{\end{equation}}
\def\ergs{\rm \ erg \, s^{-1}}
\def\deg {^\circ}

\def\cmdue {\rm \ cm^{-2}}

\def\msole {~M_{\odot}}

\begin{document}

\title{The 1998 outburst of the X--ray transient XTE J2012+381 as observed
with BeppoSAX}
\titlerunning{BeppoSAX observations of XTE J2012+381}

\author{S. Campana\inst{1} \and
L. Stella\inst{2,}\thanks{Affiliated to I.C.R.A.} \and T. Belloni \inst{1}
\and G.L. Israel\inst{2}$^{, \star}$ \and A. Santangelo\inst{3}
\and F. Frontera\inst{4,5} \and M. Orlandini\inst{5} \and D. Dal Fiume\inst{5}}

\institute{Osservatorio Astronomico di Brera, Via E. Bianchi 46,
I--23807 Merate (Lecco), Italy
\and
Osservatorio Astronomico di Roma, Via Frascati 33,
I--00040 Monteporzio Catone (Roma), Italy
\and
Istituto di Fisica Cosmica e Applicazioni all'Informatica, C.N.R.,
Via La Malfa 153, I--90146 Palermo, Italy
\and
Universit\`a di Ferrara, Dipartimento di fisica, Via Paradiso 12,
I--44100 Ferrara, Italy
\and
Istituto Tecnologie e Studio Radiazioni Extraterrestri, C.N.R.,
Via Gobetti 101, I--40129 Bologna, Italy
}

\date{Received  / Accepted }
\offprints{S. Campana: campana@mera\-te.mi.astro.it}

\abstract
{
We report on the results of a series of X--ray observations of the transient
black hole candidate XTE J2012+381 during the 1998 outburst performed with
the BeppoSAX satellite.
The observed broad-band energy spectrum can be described with the
superposition of an absorbed disk black body, an iron line plus a high
energy component, modelled with either a power law or a Comptonisation tail.
The source showed pronounced spectral variability between our five 
observations. While
the soft component in the spectrum remained almost unchanged throughout our
campaign, we detected a hard spectral tail which extended to
200 keV in the first two observations, but became barely
detectable up to 50 keV in the following two.
A further re-hardening is observed in the final observation.
The transition from a hard to a soft and then back to a hard state occurred
around an unabsorbed 0.1--200 keV luminosity of $10^{38}\ergs$ (at 10 kpc).
This indicates that state transitions in XTE 2012+281 are probably not driven
only by mass accretion rate, but additional physical parameters must play
a role in the evolution of the outburst.
}
\maketitle


\section{Introduction}

\begin{figure*}[htb]
\psfig{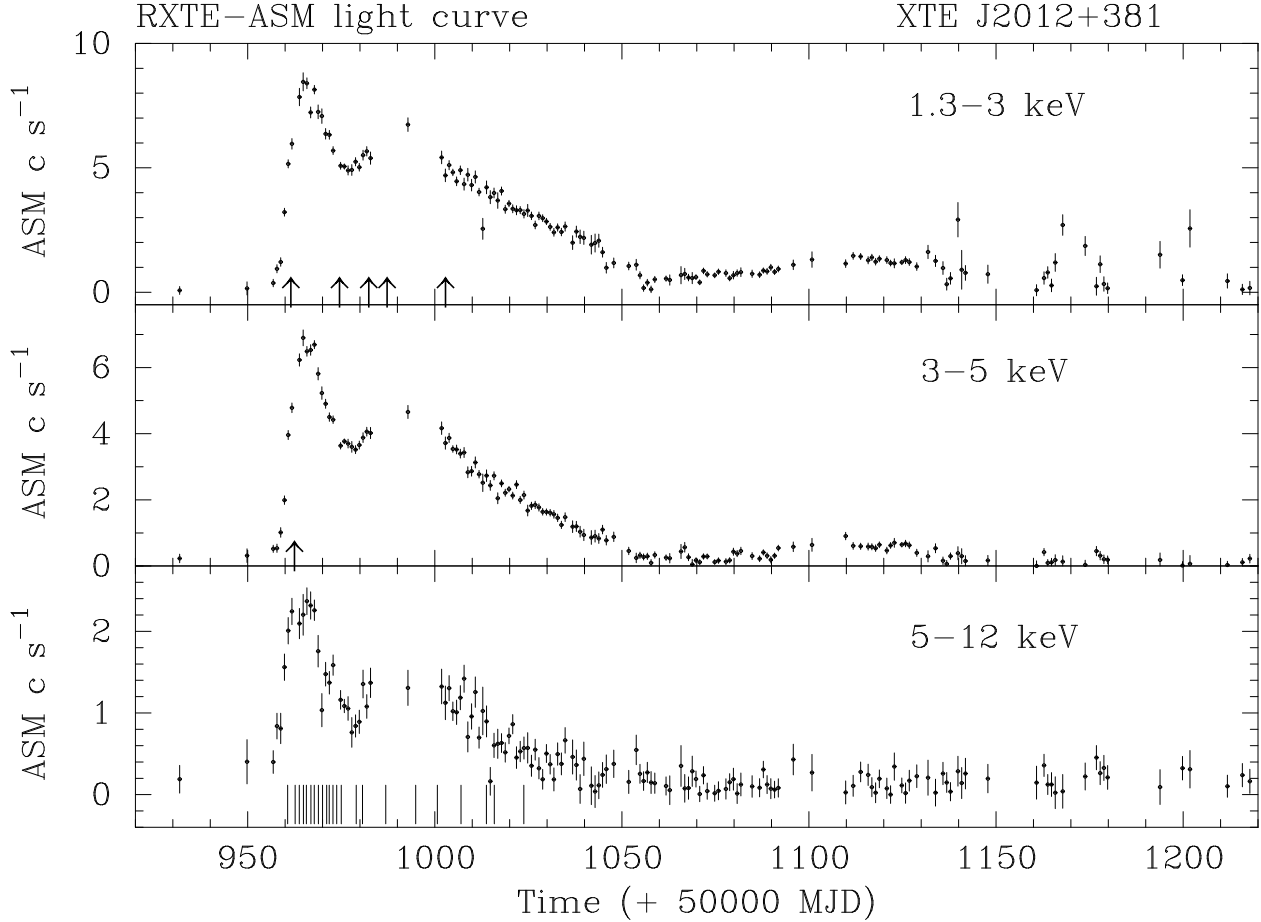}
\caption[h]{X--ray light curve of XTE J2012+381 as observed with the 
RXTE-ASM in the three energy band (1.3--3, 3--5 and 5--12 keV). The arrows 
in the upper panel mark the epochs of the five BeppoSAX observations; the
arrow in the middle panel marks the epoch of the ASCA observation whereas the
bars at the bottom of the lower panel show the epochs of the RXTE pointed
observations.} 
\label{lc}
\end{figure*}

The transient X--ray source XTE J2012+381 was discovered with the Rossi
X--ray Timing Explorer All Sky Monitor (RXTE-ASM) on May 24, 1998 at a level
of 23 mCrab (2--12 keV; Remillard, Levine \& Wood 1998; see
Fig. \ref{lc}). Within 3 days the source raised to an average level of
88 mCrab (May 27, 1998), reaching values as high as 110 mCrab (2--10
keV, RXTE Proportional Counter Array - PCA; Marshall \& Strohmayer 1998).  
An ASCA observation on May 29, 1998 (MJD 50962) show\-ed 
the source at a level 150 mCrab (2--10 keV; White et al. 1998). 
The ASCA Gas Imaging Spectrometer 
0.5--10 keV spectrum could be well described by the superposition of a
multicolor disk black body and a power law model.  
The temperature of the innermost disk radius was $k\,T=0.76\pm 0.01$ keV, 
the power law photon index $\Gamma=2.9\pm0.1$ and the absorption column density
$N_H=(1.29\pm0.03)\times 10^{22}\cmdue$ (White et al. 1998). 
The presence of a soft thermal component (equivalent temperature $\sim 1$ keV) plus 
a hard power law is considered to be a characteristic feature of black hole
candidates (e.g. Tanaka \& Shibazaki 1996). 
The outburst evolution as observed by the RXTE-ASM is shown in Fig. \ref{lc}.
An extensive analysis of the available RXTE PCA pointings was carried out 
by Vasiliev, Trudolyubov \& Revnivtsev (2000). 

Very Large Array (VLA) observations on 31 May, 1998 led to the
identification of a radio source (Hjellming, Rupen \& Mioduszewski 1998a).
The radio source was located at R.A.=20$^{\rm h}$12$^{\rm m}$37$^{\rm s}.67$
and Dec.=+38$\deg$11$^{\rm '}$01$^{\rm ''}.2$ (equinox 2000) within the $\sim
1{\rm'}$ RXTE $90\%$ error circle (Hjellming, Rupen \& Mioduszewski 1998b).

Despite the large column density (converting to a V-band extinction of about
7 mag) inferred from the X--ray spectrum, the large flux variations
characteristic of X--ray novae made possible the identification of a faint
red (R=20.1; V=21.3) counterpart during the X--ray source outburst, at
a position consistent with the location of the radio counterpart (Hynes \&
Roche 1998; Hynes et al. 1999).

Here we report on an observation campaign carried out with the Italian-Dutch 
satellite BeppoSAX (Boella et al. 1997a), aimed at studying in detail the outburst 
of XTE J2012+381. Five observations
were carried out starting from May 28, 1998  
through July 8, 1998. The X--ray spectra and light curves are discussed in
Section 2. Section 3 is dedicated to the discussion of the results; our 
conclusions are summarised in Section 4. 

\begin{table*}
\caption{Summary of BeppoSAX observations.}
\begin{tabular}{cc|cc|cc|cc|cc}
Obs.& Start   & LECS      & LECS       & MECS      & MECS       & HPGSPC    & HPGSPC    & PDS       & PDS         \\
num.& time    & Exp. time & Count rate & Exp. time & Count rate & Exp. time & Count rate& Exp. time & Count rate  \\
    & MJD     & (s)       & c s$^{-1}$ & (s)       & c s$^{-1}$ & (s)       & c s$^{-1}$& (s)       & c s$^{-1}$  \\
\hline
1   &50961.554& 11540     & 24.4       & 23641     & 19.6       & 11544	    & 11.2      & 11973     & \ 2.7       \\
2   &50974.535& \ 4151    & 20.1       & \ 7916    & 17.5       & \ 2769    &\ 1.4      & \ 2396    & \ 1.4       \\
3   &50982.365& 16495     & 22.9       & 27339     & 20.0       & 11160     &\ 0.6      & 10281     & \ 0.2       \\
4   &50987.250& 19388     & 26.7       & 31500     & 22.2       & --        & --        & 13282     & $<0.2^*$    \\
5   &51002.854& 15214     & 19.7       & 27142     & 17.0       & --        & --        & 12037     & 0.3         \\
\end{tabular}

\noindent Count rates are in the energy bands used in the spectral analysis (see text).

\noindent $^*$ $3\,\sigma$ upper limit.
\label{date}
\end{table*}

\section{Data analysis}

We analysed the data from all the BeppoSAX narrow field instruments: the Low Energy 
Concentrator Spectrometer (LECS; 0.1--10 keV, Parmar et al. 1997), 
the Medium Energy  Concentrator Spectrometer (MECS; 1.3--10 keV, 
Boella et al. 1997b), the High Pressure Gas Scintillation Proportional Counter 
(HPGSPC; 4--120 keV, Manzo et al. 1997) and the Phoswich Detection System (PDS; 
15--300 keV, Frontera et al. 1997). At the time of the observations, only two 
of the three MECS units were operating. LECS data were collected only
during satellite night-time leading to substantially shorter exposure times
compared to the MECS.
The HPGSPC and PDS collimators were alternatively rocked on and off the 
source to monitor the background, effectively halving their  
exposure times. 
During an additional BeppoSAX observation that took place on MJD 50967 the MECS
detectors were turned off and the satellite experienced several drifts
resulting in different source locations in the focal plane. We excluded this  
observation from further analysis. 

\begin{table*}[htb]
\caption{Spectral parameters (disk black body + power law + Gaussian line) 
the BeppoSAX observations.}
\begin{tabular}{lcccccccc}
Obs. & $N_H$               & $k\,T$      &  Disk radius & Power law   & Line energy  & Line width  &Line $EW$& $\chi^2$  \\
     &($10^{22}$ cm$^{-2}$)& (keV)       &  (km)$^{\#}$ &photon index &  (keV)       & (keV)       & (eV)    & red. (dof)\\
\hline
1    &$1.30\pm0.02$        &$0.75\pm0.01$&$33.3\pm0.7$  &$2.22\pm0.05$&$5.95\pm0.15$ &$1.13\pm0.15$& 343     & 1.09 (244)\\
2    &$1.24\pm0.02$        &$0.72\pm0.01$&$38.2\pm1.3$  &$2.12\pm0.28$&$5.92\pm1.50$ &$1.26\pm0.71$& 409     & 1.16 (172)\\
3    &$1.32\pm0.06$        &$0.73\pm0.01$&$38.5\pm0.9$  &$3.07\pm0.29$&$5.89\pm1.90$ &$0.88\pm1.18$& 145     & 0.95 (197)\\
4$^\ddag$&$1.27\pm0.03$    &$0.76\pm0.01$&$37.9\pm0.7$  &$3.33\pm1.40$& 6.00 (fixed) &$1.60\pm0.35$& 297     & 1.15 (152)\\
5    &$1.25\pm0.02$        &$0.72\pm0.01$&$38.9\pm1.1$  &$1.82\pm0.46$&$5.84\pm0.88$ &$1.10\pm0.73$& 320     & 0.98 (150)\\
\hline
\end{tabular}

\noindent All the errors are at $90\%$ confidence level for one interesting 
parameter. 

\noindent $^{\#}$ We assumed a distance of 10 kpc and an inclination of
0 degrees. 

\noindent $^\ddag$ The spectral modelling of this observation turned to be
extremely problematic. Only by fixing the value of the iron line (to 6.0 keV)
a reasonably good fit can be achieved.  

\label{spectral}
\end{table*}

\subsection{Spectral analysis}

The LECS and MECS data were extracted within a radius of $8'$ centered on the
source position. Background subtraction was applied even if it influence is
only marginal (it comprises $99.9\%$ of the total flux).
We rebinned the LECS and MECS spectra using the grouping files provided 
by SAX-SDC, {\tt lecs\_5.grouping} and {\tt mecs\_5.grouping}, respectively, 
(these files allow to sample the instrument resolution with the same
number of channels, 5, at all energies).
Given the large number of counts (allowing to reach in each rebinned channel
an uncertainty around $2\%$, i.e. the accuracy level of the MECS calibration
for bright source) we consider only the MECS2 unit.
HPGSPC data were extracted using SAXDAS (hpproducts V3.0.0). PDS data were
extracted using XAS.
The HPGSPC and PDS background was estimated with standard procedures using
off-axis blank fields.  
HPGSPC and PDS data were grouped in bins which are (nearly) logarithmically
spaced.  
Given the large number of counts we carried out the spectral analysis
in the energy ranges where the instrumental responses are best known: 0.3--4
keV for the LECS, 1.8--10 keV for the MECS, 8--20 keV for
the HPGSPC and 15--200 keV for the PDS.
We used the publicly available calibration files of January 2000 and XSPEC
11.1.0. We added a systematic error of $2\%$ to account for systematic
uncertainties in the calibrations.
This is in line with previous studies of bright galactic sources
(e.g. Oosterbroek et al. 2001).
A variable normalisation factor was used in the spectral analysis to account
for the mismatch in the (small) absolute flux calibration between the
different BeppoSAX instruments (Fiore, Guainazzi \& Grandi 1999).

\begin{figure}[htb]
\psfig{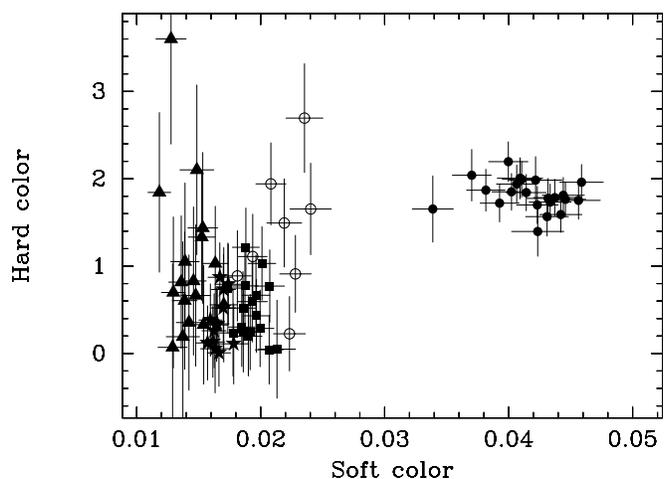}
\caption[h]{Color-color digram for the five BeppoSAX observations of XTE
J2012+381. Filled dots, empty dots, filled squares, filled stars and filled 
triangles mark the colors of the 1st, 2nd, 3rd, 4th and 5th observations, 
respectively. Soft color data are obtained as the ratio of 6--10 keV to 2--6 keV 
(background subtracted) counts in the MECS; hard color data as the ratio of 
15--50 keV to 6--10 keV (background subtracted) counts in the PDS and MECS, 
respectively. Counts are evaluated in a 1,000 s time bin.}
\label{color}
\end{figure}

As can be observed from Table \ref{date}, the PDS and HPGSPC count rates 
changed considerably across the observational campaign with a drop by a factor
of larger than 10. At lower energies the count rates underwent significant, though
considerably smaller, variations. A simple and model-independent way of
looking at these variations is to construct a color-color diagram. We considered
three energy bands: Soft (S, 2--6 keV, MECS), Medium (M, 6--10 keV, MECS) and 
Hard (H, 15--50 keV, PDS). We defined a soft color obtained as the ratio  
M/S and a hard color as H/M. The color-color diagram for the entire set of 
observations is shown in Fig. \ref{color}. The first observation (filled dots)
is clearly separated from the others both in terms of soft and hard colors;  
also the second observation (open dots) is harder than the others. The
remaining three observations lay all in the same region with low soft color.

We first tried single component models to fit the data, but all of them provided
poor fits (reduced $\chi^2 \gsim 20$).
We tried different two-component models made by a disk black body component
plus a power law, a cut-off power law or a Comptonisation tail. We first focus
on the first model and discuss in the following the other two.  

\begin{figure}[htb]
\psfig{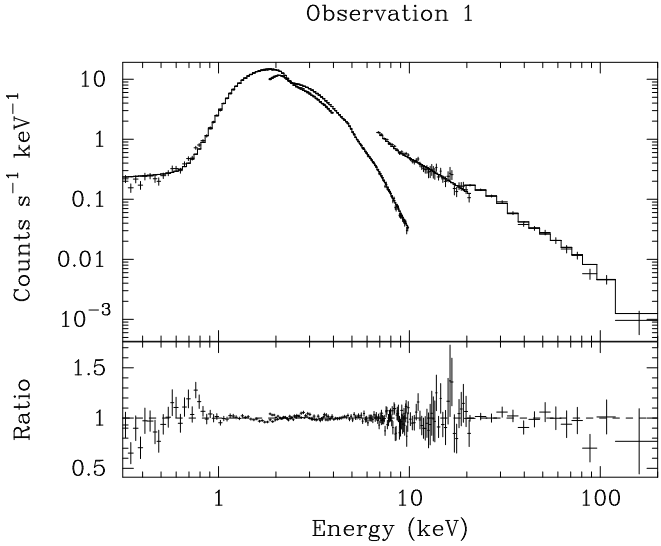}
\caption[h]{X--ray spectrum of the first BeppoSAX observation of XTE
J2012+381 with the disk black body power law model. Residuals from the fit are 
shown in the lower panel.}
\label{spe1}
\end{figure}

As reported by White et al. (1998) using ASCA data and by Vasiliev et al. 
(2000) using RXTE data, the standard model for black hole candidates 
consisting of a disk black body (Mitsuda et al. 1984) plus a power law 
and a Gaussian iron line provides an adequate fit to those data. As reported 
in Table \ref{spectral} (see also Fig. \ref{spe1}), this model provided also an 
adequate fit to the entire BeppoSAX dataset. Absorbed and unabsorbed fluxes in
different energy bands are reported in Table \ref{flux}.  

\begin{figure*}[htb]
\psfig{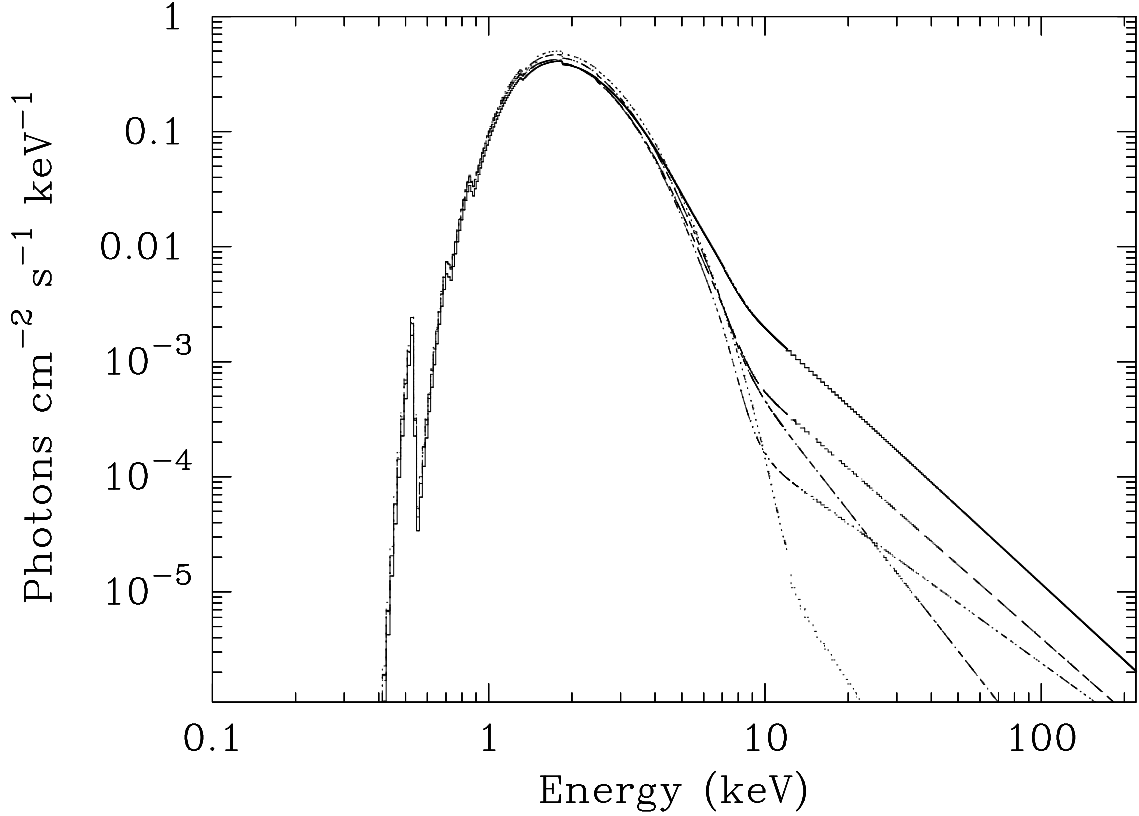}
\caption[h]{Absorbed model spectra of the five BeppoSAX observations of
XTE J2012+381, assuming a disk black body plus power law model. Spectra
from different observations are indicated by different line styles (1st:
continuous; 2nd: dashed; 3rd: dot-dot-dot-dashed; 4th: 
dotted; 5th: dot-dashed style). Note that for the fifth observation the 
extrapolation at high energies is rather uncertain.}
\label{model}
\end{figure*}

\begin{figure*}[htb]
\psfig{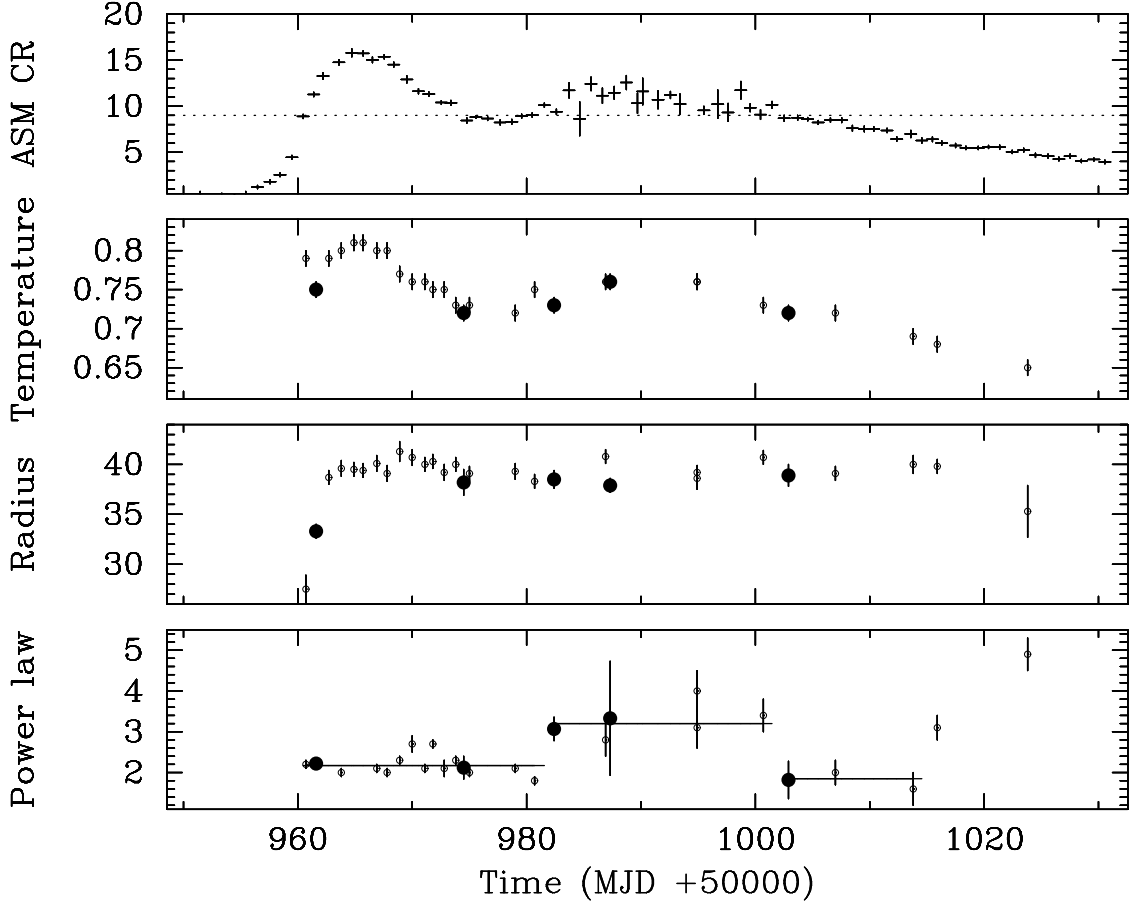}
\caption[h]{Evolution of the light curve and spectral parameter. From top to bottom, 
count rate, temperature and radius of the disk black body component, and power law 
photon index. 
In the first panel a rough indication of the flux at which spectral changes
occur is drawn with a dotted line.
In the lowest panel a fit with a constant is overlaid for the three periods 
discussed in the text. Filled dots indicate values derived from
BeppoSAX observations. Small open dots indicate values obtained with RXTE
(Vasiliev et al. 2000).} 
\label{evol}
\end{figure*}

This model parametrisation allows us to monitor the spectral evolution of 
XTE 2012+381 (see Fig. \ref{model}) and to compare the results with
observations from other instruments.
For the column density we limit our analysis to the BeppoSAX data and
model it with the new XSPEC model {\tt TBABS} (which includes updates to the
photoionisation cross section, revised abundances of the interstellar
medium, and presence of interstellar grains and the H2 molecule; Wilms, Allen
\& McCray 2000). 
This new model gives significantly better fits then the usual {\tt WABS} 
model and a slightly lower column density value. Across the BeppoSAX
observational campaign the value of the column density remains unchanged
with a mean value of $(1.27\pm0.02)\times 10^{22}\cmdue$. This is consistent
with the Galactic value of $1.2\times 10^{22}\cmdue$ (Dickey \& Lockman 1990). 
In the analysis of the other spectral parameters we include also the values
derived from RXTE observations (Fig. \ref{evol}).
The parameters of the soft spectral component display small but significant 
variations: the temperature follows closely the flux evolution whereas the
radius shows a steep increase from the first observation and then settles to a
constant value of $R=39.5\pm0.3$ km (see Fig. \ref{evol}).
The iron line properties remain constant with a mean centroid $E=5.9\pm0.3$ keV and
a mean line width ($\sigma$) $W=1.2\pm0.2$ keV (these values are based
only on BeppoSAX analysis  
since Vasiliev et al. keep the iron line energy fixed at 6.4 keV).
Broad and redshifted iron lines are rare among X--ray binaries. This is
potentially extremely interesting however we note that the power law and the
disk black body components become comparable just around 6 keV. We speculate
that part of the line width is to ascribe to a non-perfect matching of the two 
components. The same problem is also present in the RXTE data.

The slope of the power law component remained constant in the
first part of the outburst up to MJD 50981 (with $\Gamma=2.2\pm0.1$),
steepened in the second part of the outburst ($\Gamma=3.2\pm0.3$; see
Fig. \ref{evol}) and it hardened again around MJD 51000. 
This change occurs simultaneously with the reflare in the outburst light
curve. 
As a result of this, the extrapolated flux at high energies ($>20$ keV) drops
smoothly by a factor of $\gsim 800$ (see Table \ref{flux}). This value is
based on the hard flux derived from spectral fits (and therefore for the third 
and fourth observations is extrapolated). If we consider the count rate
variation the drop is by a factor of $>10$. 

In the last BeppoSAX observation a re-hardening is observed (see
Figs. \ref{model}, \ref{evol}).
In this observation the power law is as hard as in the first two
observations and also RXTE observations contiguous (in time) provide
similar hard power laws ($\Gamma=1.9\pm0.4$; see Fig. \ref{evol}). It is
interesting to note that the flux level at 
which this re-hardening occurs is just below a threshold of $\sim 9$ ASM c
s$^{-1}$ (see Fig. \ref{model}), which is also the threshold at which the
the first flare ends and the reflare starts. 

We also tried different models such as {\tt PEXRAV} (consisting of an 
exponentially cut-off power law plus the reflected component from neutral 
material; Magdziarz \& Zdziarski 1995), or Comptonisation models (e.g. {\tt 
COMPTT}, Titarchuk 1994), together with a disk black body and an iron line.
The {\tt PEXRAV} model provides an equally good fit to the first observation 
($\chi^2_{\rm red}= 1.09$) but XSPEC fails to converge for the second one.
Moreover, the reflected fraction is very low. For these reasons we do not
consider this model. 
The {\tt COMPTT} model gives $\chi^2_{\rm red}= 1.09,\, 1.12,\, 0.92,\, 
1.03,\, 0.98$ for the five observations, respectively.
The {\tt COMPTT} model provides a better fit with respect to the power law
model on the entire dataset at $99.4\%$ confidence level (estimated by means
of an F-test). We decide to discuss first the disk black body plus power law
model in order to include also the RXTE data analysed by Vasiliev et
al. (2000). 
In Table \ref{compto} we report the parameters of the {\tt COMPTT} model only,
with the column density, soft component and iron line remaining almost
the same. A clear trend can also be observed in this spectral decomposition
following closely the source flux and with the hardest observations having the
smallest input soft photon (Wien) temperature ($T_0$) and the highest plasma
temperature $T_{\rm C}$.

We also considered a model with a cut-off power law. In this case we have
problems in disentangling the power law photon index and the cut-off energy.
What occurred is that the power law becomes steeper and steeper and the cut-off 
energy lower and lower (e.g. for the second observation we obtained a power law
photon index --0.5 and a cut-off energy of 10 keV). In order to monitor the 
cut-off energy in the different spectral states we decided to fix 
the power law index to 2 and to derive the cut-off energy. This energy turned
out to decrease (as expected) with the softening of the spectrum (see Table
\ref{compto}). 

\begin{table}[htb]
\caption{Parameters of the Comptonisation and cut-off power law models.}
\begin{tabular}{lccc|c}
Obs. & $T_0$        & $T_{\rm C}$  & $\tau$        & $E_{\rm cutoff}$\\  
     & (eV)         &    (keV)     &               &  (keV)\\
\hline
1    &  8           &   152.5      & 0.09          & 205 \\
2    &  62          &    96.6      & 0.22 (fixed)  & 134 \\
3    &  26          &    45.4      & 0.17          & 8.8\\
4    &  46          &    14.0      & 0.08          & 8.0\\
5    &  31          &   169.5      & 0.22          & $>80$\\
\end{tabular}

\noindent Fitting the second observation proved problematic and we have to 
fixed the value of at least one parameter to make the fit converge. Given the 
similarities between the second and the fifth observation, we decided to fix
the optical depth $\tau$ in the {\tt COMPTT} model to 0.22. The fit in any
case was highly unstable and we cation the reader on the results obtained.
\label{compto}
\end{table}

\begin{table*}[htb]
\caption{X--ray (absorbed) fluxes for the BeppoSAX observations with power law 
fit.}
\begin{tabular}{ccccccc|c}
Obs.&0.1--1 keV      & 2--10 keV     & 20--200 keV    & 0.1--200 keV  & DBB flux 2--10 keV  & PL flux 2--10 keV& 0.1--200 keV UNABS\\
    &($10^{-11}$ cgs)&($10^{-9}$ cgs)&($10^{-11}$ cgs)&($10^{-9}$ cgs)& ($10^{-9}$ cgs)     & ($10^{-9}$ cgs)  & ($10^{-8}$ cgs)\\
\hline
1   & 1.39           & 2.47          & 48.25          & 3.96          &  1.92               &   0.52           &1.08\\
2   & 1.63           & 2.10          & 15.44          & 3.16          &  1.96               &   0.14           &0.90\\
3   & 1.66           & 2.39          &\ 2.80          & 3.38          &  2.17               &   0.22           &1.77\\
4   & 1.80           & 2.66          &\ 0.24          & 3.67          &  2.60               &   0.06           &1.30\\
5   & 1.62           & 2.00          &\ 8.00          & 2.94          &  1.96               &   0.04           &0.86\\
\end{tabular}
\label{flux}

\noindent Fluxes are derived from spectral fits. In particular, the 20--200
keV flux of the fourth observation is extrapolated from the model, whereas the 
PDS provided only an upper limit.
\end{table*}

\subsection{Timing analysis}

For each observation, we produce a r.m.s. normalised power spectrum of the
1.8--10 keV light curve to study 
the timing properties of the system. Both MECS units are used in this case. 
Power spectra do not show prominent noise components. No periodicities or 
quasi-periodicities are present with an upper limit of $\lsim 5\%$ over 
the $10^{-2}-1$ Hz range for the five observations discussed; this limit
increases to $\sim 60\%$ 
at $10^{-4}$ Hz. In order to better characterise the spectral states of 
XTE J2012+381 we compute the fractional r.m.s. variability (for frequencies 
higher than 1 Hz): this is $<4.7\%$ ($3\,\sigma$ level), 
$<5.4\%$, $<6.4\%$, $<8.6\%$ and $<7.6\%$ for the five MECS observations, 
respectively. 
A more detailed analysis is carried out with the RXTE light curves.
By averaging observations between MJD 50962 and MJD 50987, Vasiliev et al. (2000)
obtained a power spectrum dominated by the very low frequency noise with a 
$\nu^{-1}$ dependence. The fractional r.m.s. variability was at a level of $\sim 2\%$.
An indication of the increase of the fractional r.m.s. with 
energy was also found. No quasi-periodic oscillations were detected.
Here we reanalysed the same set of data splitting the RXTE data set into 
three periods. For the first period (between MJD 50962 and MJD 50981 encompassing 
the first two BeppoSAX observations) we derive a r.m.s. variability of $1.7\%$. For 
the 4 RXTE observations between MJD 50986 and MJD 51001 (including the
third and fourth BeppoSAX observation) the fractional r.m.s. variability
raised to $3\%$. Finally the four last RXTE observations in between MJD
51007 and MJD 51024 provided only an upper limit of $<1.2\%$. 

\subsection{Quiescent emission}

The field containing XTE J2012+381 was monitored a few times by ROSAT
when observing the reflection nebula NGC 6888. Three observations were carried 
out with the HRI detector for a total observing time of 109 ks. One observation 
was carried out with the PSPC detector for 8 ks. In none of the observations
XTE J2012+381 was detected. The tightest upper limit is provided by the three
HRI observations summed together.
Assuming the same column density observed during the 1998 outburst, the
extrapolated $3\,\sigma$ upper limit on the quiescent 0.1--2.4 keV
(unabsorbed) luminosity is $\sim 2\times 10^{33}\ergs$ for a Crab
spectrum and a distance of 10 kpc\footnote{An upper limit on the distance
$d\lsim 12$ kpc comes from the requirement that XTE J2012+381 lies within the
Galaxy. The column density value inferred from the BeppoSAX data, being
compatible with the full  galactic value, favours a distant object. In the
following we assume a value of $d=d_{10}\,10$ kpc.}). This limit is fairly
high and it is mainly due to the low energy pass band together with the high column
density. This limit alone cannot allow us to disentangle the nature of the
compact object in XTE J2012+381 (e.g. Campana \& Stella 2000).  

\section{Discussion}

BeppoSAX observations as well as other X--ray observations made possible the 
identification of an ultra-soft component ($k\,T\sim 0.7$ keV) together with a hard 
energy tail (photon index $\sim 2$; see also White et al. 1998; Vasiliev 
et al. 2000) extending up to 200 keV in the outburst X--ray spectrum of XTE J2012+381.
Spectral variability was observed with the hard power law count rate
decreasing by a factor $\gsim 10$ in the 20--200 keV band and a factor of $\gsim 800$
in flux according to the fitted spectra. These characteristics, 
together with the lack of X--ray bursts and coherent pulsations, clearly suggest 
a black hole nature of XTE J2012+381.

The absorbed 0.1--200 keV luminosity during the BeppoSAX
observations was at a level of $\sim 4\times 10^{37}\,d_{10}^2\ergs$ and showed only
modest variations ($\sim 30\%$) across the different observations. 
Given the large column density, removing the effects of the interstellar 
absorption results in a substantially higher luminosity: the 0.1--200 keV 
unabsorbed luminosity is 1.3, 1.1, 2.1, 1.6, 1.0 $\times 10^{38}\,d_{10}^2$ 
erg s$^{-1}$, however.

\subsection{Taxonomy of source states}

Different spectral states have been identified in black hole candidates (BHCs)
as their spectral and timing properties vary along with their X--ray 
luminosity (for a review see e.g. van der Klis 1995).
In the {\it low state} (LS) BHCs are characterised by a power law spectrum 
($\Gamma\sim 1.5-2$) and (sometimes) a weak disk component ($k\,T\lsim 1$ keV).
Strong rapid aperiodic variability with fractional r.m.s. 
amplitude of $20-50\%$ and a break frequency below 1 Hz is also observed.
At increasing X--ray fluxes, the source enters the {\it intermediate state} (IS), 
characterised by a steep power law ($\Gamma\sim 2-3$) 
plus a disk component. The timing properties present a broad band noise 
with $\sim 5-20\%$ r.m.s. and a break frequency higher than 1--10 Hz. 
The {\it high state} (HS) is dominated by the disk component with a weak and 
steep power law ($\Gamma\sim 2-3$). A few percent r.m.s. characterise the 
temporal variability and this is its main difference with the IS. 
Finally, a {\it very high state} (VHS) has been observed in 
a few BHCs, with a spectrum similar to the one in the intermediate state (disk 
component with $k\,T\sim 1-2$ keV and power law with $\Gamma\sim 2.5$), and with  
band limited noise at $1-15\%$ r.m.s. level. Quasi-periodic oscillations at 1--10 Hz 
are often seen (e.g. Miyamoto et al. 1991; Homan et al. 2000).
This classification scheme is one-dimensional with transitions following
LS $\to$ IS $\to$ HS $\to$ VHS for increasing mass accretion rates. 
Homan et al. (2000) analysed the RXTE data of the BHC XTE J1550--564
and noted that the simple correlation of states with accretion rate breaks
down. Given the similarity of the IS and the VHS, they 
concluded that the VHS is an instance of the IS. Based on this,
the classification scheme that they propose envisages a two-dimensional
behaviour with the count rate and the spectral hardness as phenomenological
parameters. In this framework the difference between the IS and the VHS is reduced 
to a difference in the count rates. (The two physical parameters are suggested to 
be the mass accretion rate and extent of a Comptonising corona, responsible 
for the hard component).
   
In all our observations the 
fractional r.m.s. variability is low and the disk (soft) component is strong 
comprising the great majority of the source luminosity. These properties
likely exclude that XTE 2012+381 entered the LS.
In the first two observations of XTE J2012+381 a hard power law with 
photon index $\sim 2$ (up to $\sim 200$ keV) is clearly detected (see Fig. 
\ref{model}). 
These characteristics together with the timing properties outlined above 
likely place XTE 2012+381 on a hard branch (i.e. the IS/VHS).
In the third and fourth observations XTE J2012+381 has a much softer spectrum 
(with a power law photon index $\sim 4$ barely detected in the PDS up to 
$\sim 50$ keV in the third) indicating that the source entered the HS.
This is further corroborated by the fact that in the fourth observation the 
count rate is the highest. 
Finally in the last observation the source came back with a relatively hard
power law (even if affected by a large uncertainty) and a small r.m.s.
variability, likely indicating that it returned to the hard branch.
The source count rate being much larger in the first two observations than 
in the last, in the `old' classification the first two might be ascribed to the VHS 
and the last to the IS.

\subsection{Comparison with other black hole candidates outbursts}

Finally, we note that the outburst evolution shares similar characteristics 
with well known BHC transients, even though the characteristic time\-scales
are shorter: $i)$
after the outburst peak (around May 22th 1998, MJD 50955.6) the RXTE-ASM
light curve decayed exponentially with an $e-$folding time of 16 d. This
time is 24, 31 and 30 d for A 0620--00, GS 1124--684, GS 2000+251,
respectively.  $ii)$ a secondary peak occurred $\sim 30$ later (Jun. 26th
1998, MJD 50990.6), reaching a level of $\sim 75\%$ the outburst peak.  A
secondary peak occurred 55, 60--75 and 70 d after maximum in A 0620--00,
GS 1124--684, GS 2000+251, respectively (e.g. Chen, Livio \& Gehrels
1993).  $iii)$ after the second peak the X--ray light curve decayed
exponentially with an $e-$folding time of 35 d, i.e. more slowly than
after the first peak.  An $e-$folding time of 20, 37 and 30 d was
observed in A 0620--00, GS 1124--684 and GS 2000+251, respectively.  
$iv)$ there was evidence for a possibly third broad peak
about 150 d after the outburst peak (Oct. 24th 1998, MJD 51110.6). The same
feature was detected in A 0620--00 ($\sim 200$ d after the outburst) and in GS
2000+251 ($\sim 70$ d).

Concerning the spectral evolution, the low energy portion of the spectra 
remained basically unchanged during the outburst evolution. At high energies,
the power law tail is hard and steepens later on. This spectral behaviour 
is similar to the one observed in GS 1124--68 (Nova Muscae
1991; Tanaka 1992), though (again) on a shorter temporal base: around the
outburst maximum the spectrum was hard but it was softer at the time of the
secondary outburst. In GS 1124--68 a hard tail was observed again in the
observations following the secondary maximum, about 120 d after the main peak,
similarly to what occurred in XTE 2012+381 $\sim 40$ d after the main peak.
These similarities further strenghten the identificationo f XTE 2012+381 with
a black hole candidate.

\section{Conclusions}

BeppoSAX observations that covered the 1998 outburst of XTE J2012+381  
found several similarities with known BHCs.
In particular, the X--ray spectrum could be well described by an 
ultrasoft component (disk black body) with the addition of a variable 
hard power law. 

One can derive an upper limit on the black hole mass by interpreting the inner 
disk radius from the spectral fits as the last marginally stable
orbit. The lowest  
values are obtained for the first observation with an equivalent radius 
$R=33$ km. For a disk inclination of $0\deg$, we derive $M\gsim 3.7\,d_{10}\msole$ 
($M\gsim 22\,d_{10}\msole$) for a non-rotating (maximally rotating) black hole.
Notice that the absolute value inferred for the disk radius, due to the 
approximations in the disk black body model used, is to be considered also 
a lower limit (Merloni, Fabian \& Ross 2000).

One of the main results of the present paper is the different spectral states
observed during the outburst evolution of XTE 2012+281. The column density,
the parameters of the soft component and iron line remained almost constant. On
the other hand the hard tail changed considerably across different observations. 
During the first flare (between MJD 50960 and 50982) the X--ray spectrum was
hard with a power law index $\Gamma=2.2\pm0.1$. In the second flare (between
MJD 50982 and 51002) the X--ray spectrum firstly became soft and then it
re-hardened (between MJD 51003 and 51014). At the outburst end, there is a hint
for a further softening. In passing, we note that BeppoSAX and RXTE
spectral fits are in good agreement, despite the poorer resolution of RXTE 
spectra. Moreover, our timing analysis put in evidence a higher r.m.s. 
variability in the interval during which the spectrum of XTE 2012+281 is soft.

The transition from a hard to a soft and then back to a hard state occurs
around the same count rate in the RXTE-ASM around ($\sim 9$ c s$^{-1}$).
Clearly for such an absorbed source the ASM cannot provide a fair estimate
of the source luminosity. Comparing BeppoSAX (unabsorbed) luminosities derived
from the spectral decomposition we see that this threshold is
around $10^{-8}\ergs\cmdue$, i.e. $10^{38}\,d^2_{\rm 10 kpc}\ergs$.
Above this flux level, during the first flare the spectrum of XTE 2012+281 is
hard, whereas during the second flare above the same flux level the source
spectrum is soft. Simultaneously, the r.m.s. fractional variability increases
only slightly from $\sim 2\%$ to $\sim 3\%$. 

This likely indicate that state transitions in XTE 2012+281 are not driven
only by mass accretion rate and at least one other parameter is needed. In
particular, the soft component remains almost stable during the BeppoSAX and
RXTE observational campaigns (within a factor of a few) whereas the hard
component varied considerably. In this regard we note that spectral parameters 
such as the plasma temperature $T_{\rm C}$ or the high energy cut-off 
vary considerably across the transition strengthening the idea that the second 
parameter involved in state transitions is related to a hot corona.

\begin{acknowledgements}
This research has made use of SAXDAS linearised and cleaned event
files (Rev.2.0) produced at the BeppoSAX Science Data Center.
This work was partially supported through ASI grants. J. Homan is acknowledged 
for useful comments. TB thanks the Cariplo Foundation for support.
\end{acknowledgements}

\end{document}